\documentclass[letterpaper,10pt]{article}

\usepackage{amssymb}
\usepackage{amsmath}
\usepackage{graphicx}
\usepackage{psfrag}

\newcommand{\dd}{\textrm{d}}

\newcommand{\hypergf}{{}_{2}F_{1}}

\title{A conditionally exactly solvable generalization of the potential step}

\author{A.\ L\'opez-Ortega \\
Departamento de F\'{\i}sica.\\ 
Escuela Superior de F\'{\i}sica y Matem\'aticas. \\
Instituto Polit\'ecnico Nacional. \\
Unidad Profesional Adolfo L\'opez Mateos. Edificio 9. \\
M\'exico, D.\ F., M\'exico. \\
C.\ P.\ 07738 \\
email: alopezo@ipn.mx }

\begin{document}

\maketitle

\begin{abstract}

Motivated by the interest in non-relativistic quantum mechanics for determining exact solutions to the Schr\"odinger equation we give two potentials that are conditionally exactly solvable. The two potentials are partner potentials and we obtain that each linearly independent solution of the Schr\"odinger equation includes two  hypergeometric functions. Furthermore we calculate their reflection and transmission amplitudes. Finally we discuss some additional properties of these potentials.



\end{abstract}

\section{Introduction}
\label{s: Introduction}

In physics a research line is the search and detailed study of systems for which we can solve exactly their equations of motion. We expect that the exactly solvable systems play a significant role since they are useful approximations to more complex systems and they allow us to make a more detailed analysis of the physical phenomena. Furthermore using these exactly solvable systems we can infer some details about the behavior of more complicated physical systems.  In non-relativistic quantum mechanics the search of potentials for which we can solve exactly the Schr\"odinger equation is widely studied. See \cite{Schrodinger}--\cite{Bhattacharjie} for an incomplete list of references.

Several methods have been used to find exact solutions to the Schr\"odinger equation. We mention the factorization method \cite{Schrodinger}, \cite{Schrodinger2}, \cite{Infeld-Hull} and the method based on supersymmetric quantum mechanics (SUSYQM) \cite{Witten-susy}--\cite{Dutt-ajp-1988}. Other widely used procedures are based on the Darboux transformation \cite{Darboux}, \cite{Bagrov} and on the point canonical transformations \cite{Bhattacharjie}. 

Recently we solve exactly the Schr\"odinger equations for the partner potentials \cite{ALO-2015-I}
\begin{equation} \label{e: fractional power potential}
 \hat{V}_\pm = \frac{m^2}{x} \pm \frac{m}{2 x^{3/2}},
\end{equation} 
where $m$ is a constant. In Ref.\ \cite{ALO-2015-I} we show that each of their linearly independent solutions includes a sum of two confluent hypergeometric functions. More recently, for $\tilde{V}_0$ a constant, in Ref.\  \cite{Ishkhanyan-1} it is showed that for the inverse square root potential 
\begin{equation} \label{e: square root potential}
 V_{1/2} = \frac{\tilde{V}_0}{\sqrt{x}}, 
\end{equation} 
its exact solutions have a similar mathematical form to those found previously in Ref.\  \cite{ALO-2015-I}, that is, each linearly independent solution is given by a linear combination of two confluent hypergeometric functions. Furthermore in Ref.\  \cite{Ishkhanyan-2} it is showed that for the sum of the potentials (\ref{e: fractional power potential}) and (\ref{e: square root potential}) it is possible to solve exactly the Schr\"odinger equation. In this case, in a similar way to the potentials  (\ref{e: fractional power potential}) and (\ref{e: square root potential}), each linearly independent solution is a sum with non-constant prefactors of two confluent hypergeometric functions.

Based on the method of Ref.\ \cite{ALO-2015-I} in this work we present two potentials for which we solve exactly their Schr\"odinger equations when the parameters satisfy a condition, thus the potentials that we study are conditionally exactly solvable (CES) in the broad sense of the concept recently introduced in Ref.\ \cite{Ishkhanyan-2}. Furthermore for each potential the linearly independent solutions of the Schr\"odinger equation include two hypergeometric functions, hence our results extend those of Refs.\ \cite{ALO-2015-I}, \cite{Ishkhanyan-1}, \cite{Ishkhanyan-2}. The  two potentials are partner potentials in the standard language of the SUSYQM. We see that the shapes of the potentials remind us to the step potential (see Figs.\ \ref{figure1} and \ref{figure3} below), thus we believe that the potentials can be useful to study scattering problems where the step-like potentials are useful models \cite{Flugge}. 

In what follows we expound in detail the method used to find the exact solutions, but here we notice that it works simultaneously with the two Schr\"odinger equations of the partner potentials \cite{ALO-2015-I}. As far as we can see these two potentials do not appear in the references that enumerate the solvable potentials that are previously known \cite{Cooper:1994eh}--\cite{Dutt-ajp-1988}, \cite{Khare-scattering}--\cite{Dutt-CES}.

We organize this paper as follows. In Sect.\ \ref{s: method} we present the partner potentials that we study throughout this work and we explore some of their properties. Furthermore we describe the method used to find the exact solutions to the Schr\"odinger equations of these potentials. In Sect.\ \ref{s: verification}  we give the solutions in explicit form and we verify that the functions found in the previous section are solutions of the Schr\"odinger equations. In Sect.\ \ref{s: scattering} we use the solutions previously found to calculate the reflection and transmission amplitudes of the studied potentials. We also determine their quasinormal frequencies. Finally in Sect.\ \ref{s: remarks} we discuss some facts on these partner potentials.

\section{Solution method}
\label{s: method}

The purpose of this work is to show that the Schr\"odinger equation\footnote{In contrast to the common usage and for simplifying some of the following mathematical expressions, we write the energy $E$ as $\omega^2$, that is, $E = \omega^2$. } 
\begin{equation} \label{e: Schrodinger equation}
\frac{\dd^{2} Z}{\dd x^{2}} + \omega^{2}  Z = V Z,
\end{equation}
is solvable for the partner potentials
\begin{eqnarray} \label{e: potentials}
V_\pm (x,m) = m^2 \frac{ e^x}{e^x + 1} \mp \frac{m}{2} \frac{e^{x/2}}{(e^x + 1)^{3/2}} =  W^2 \pm \frac{\dd W}{\dd x}, 
\end{eqnarray} 
where $m$ is a constant, the variable $x \in (-\infty, +\infty)$, and the superpotential $W$ is equal to\footnote{The Schr\"odinger equations for the partner potentials derived from the superpotential $\hat{W} = -m e^{x/2}(\mathcal{A} e^x + \mathcal{B})^{-1/2}$ with $\mathcal{A} > 0$, $\mathcal{B} > 0$ reduce to those of the potentials for the superpotential (\ref{e: superpotential}) if we make the change of variable $y = \ln \left( \mathcal{A}/ \mathcal{B} \right) + x$ and we change the parameter $m$ by $m / \mathcal{A}^{1/2}$. }
\begin{equation} \label{e: superpotential}
 W (x,m) = - m \sqrt{\frac{e^x}{e^x+1}} = - \frac{m}{(1+e^{-x})^{1/2} }.
\end{equation} 

\begin{figure}[th]
\begin{center}
\includegraphics[scale=.9,clip=true]{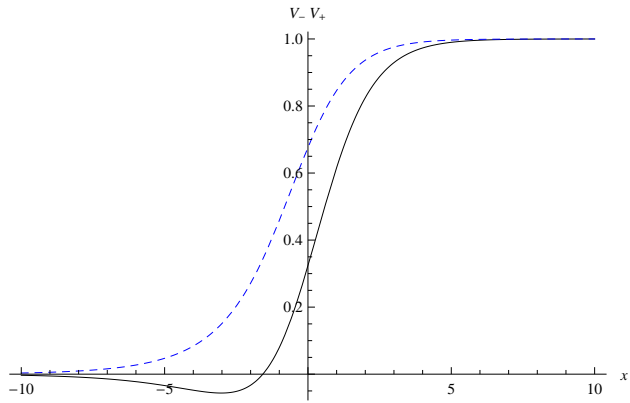}
\caption{Plots of the potentials $V_+$ (solid line) and $V_-$ (broken line)  for $m=1$. \label{figure1}} 
\end{center}
\end{figure}

The potentials (\ref{e: potentials}) behave as
\begin{equation}
 \lim_{x \to +\infty} V_\pm = m^2 ,
\end{equation} 
and for $m>0$ 
\begin{equation} \label{e: limit minus infinity potential}
 \lim_{x \to -\infty} V_+ = 0^-, \qquad \qquad \lim_{x \to -\infty} V_- = 0^+,
\end{equation} 
where $0^-$ ($0^+$) means that the potential goes to zero taking negative (positive) values. For $m < 0$, the potentials $V_+$ and $V_-$ change their places in the formulas (\ref{e: limit minus infinity potential}). It is convenient to notice that as $x \to + \infty$ the potentials $V_\pm$ go to $m^2$ and the subleading terms are of the form $\exp(-x)$, whereas they decay exponentially to zero as $x \to - \infty$. We plot the potentials (\ref{e: potentials}) in Fig.\ \ref{figure1}. Furthermore from the shape of the potentials $V_\pm$ we expect that their spectra are continuous \cite{Flugge}, \cite{Dterhaar}.

\begin{figure}[th]
\begin{center}
\includegraphics[scale=.9,clip=true]{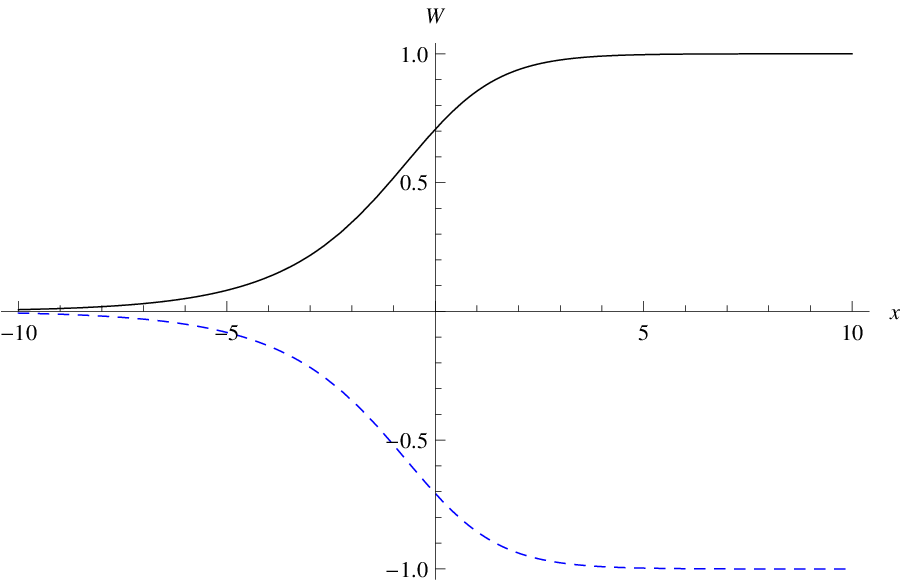}
\caption{Plots of the superpotential $W$  for $m=1$ (broken line) and for $m=-1$ (solid line). \label{figure2}} 
\end{center}
\end{figure}

For the superpotential (\ref{e: superpotential}) we obtain 
\begin{equation} \label{e: superpotential + limit}
 W_+=\lim_{x \to +\infty} W = -m,  \qquad  W_-=\lim_{x \to -\infty} W = 0. 
\end{equation} 
As $x \to - \infty$, for $m>0$ the superpotential $W$ goes to zero taking negative values, whereas for $m<0$ it goes to zero taking positive values. Since the exponential function satisfies $e^x > 0$ for $x \in (-\infty, +\infty)$ we point out that the superpotential (\ref{e: superpotential}) does not cross the $x$ axis, hence it is positive for $m < 0$ and negative for $m > 0$. We plot the superpotential (\ref{e: superpotential}) in Fig.\ \ref{figure2}. In what follows we assume that $m>0$.

Taking into account that $ e^{x} = e^{x/2} / e^{-x/2}$ we obtain that the superpotential (\ref{e: superpotential}) becomes
\begin{equation}
 W = - \mu (1 + \tanh(x/2))^{1/2},
\end{equation} 
with $\mu = m / \sqrt{2}$, and the potentials (\ref{e: potentials}) transform into
\begin{equation} \label{e: potentials hyperbolic}
 V_\pm = \mu^2  (1 + \tanh(x/2)) \mp \frac{\mu}{4} \frac{(1 - \tanh(x/2))^{1/2}}{\cosh(x/2)} .
\end{equation} 
Comparing with the potential step \cite{Flugge}
\begin{equation} \label{e: potential step}
 V_S (x) = \frac{V_0}{2} \left(1 + \tanh \left(\frac{x}{2 \alpha} \right)\right) ,
\end{equation} 
where $V_0$ and $\alpha$ are constants, we see that the potentials (\ref{e: potentials hyperbolic}), and therefore the potentials (\ref{e: potentials}), are a generalization of the potential step \cite{Flugge}. It is convenient to notice that the potential step can not be obtained as a limit of the partner potentials $V_\pm$.

Among the solvable potentials there are several examples for which it is possible to find the exact solutions only when the parameters satisfy some constraints. These potentials are called conditionally exactly solvable (CES) potentials \cite{Bagchi-book}, \cite{Ishkhanyan-2}, \cite{Souza Dutra}--\cite{Dutt-CES}. The algebraic form of some known CES potentials recall us to the potentials (\ref{e: potentials}), for example, the CES potential \cite{Bagchi-book}, \cite{Dutt-CES}
\begin{equation}
 V_I (x) =  \frac{a_I}{1 + e^{-2 x}} - \frac{b_I}{(1 + e^{-2 x})^{1/2}} - \frac{3}{4(1 + e^{-2 x})^2} ,
\end{equation} 
whose superpotential takes the form 
\begin{equation}
 W_I (x) = \frac{p_I}{(1 + e^{-2 x})^{1/2}} - \frac{1}{2 (1 + e^{-2 x})} - \sqrt{\epsilon_0},
\end{equation} 
where the constants $a_I$, $b_I$, $p_I$, and $\epsilon_0$ satisfy some constraints \cite{Bagchi-book}, \cite{Dutt-CES}. Another example is the CES potential  \cite{Bagchi-book}, \cite{Dutt-CES}
\begin{equation}
 V_{II} (x) = \frac{ a_{II} }{1 + e^{-2 x}} - \frac{ b_{II} e^{-x} }{ (1 + e^{-2 x})^{1/2} }  - \frac{3}{4 (1 + e^{-2 x})^2  },
\end{equation} 
with superpotential 
\begin{equation}
 W_{II} (x) = p_{II} + \frac{1}{2 (1 + e^{2 x})} - \frac{q_{II}}{(1 + e^{2 x})^{1/2}} .
\end{equation} 
In a similar way to the previous CES potential, the constants $a_{II}$, $b_{II}$, $p_{II}$, and $q_{II}$ satisfy some constraints \cite{Bagchi-book}, \cite{Dutt-CES}. 

Owing to the constants multiplying the factors $ e^x / (e^x + 1)$ ($m^2$) and $e^{x/2}/(e^x + 1)^{3/2}$ ($\pm m/2$) satisfy 
\begin{equation}
 \frac{1}{4} m^2 - \left( \pm \frac{m}{2} \right)^2 = 0,
\end{equation} 
we find that the potentials (\ref{e: potentials}) are CES in the sense of Ref.\  \cite{Ishkhanyan-2}, that is, we must understand that the parameters multiplying the factors  $ e^x / (e^x + 1)$ and $e^{x/2}/(e^x + 1)^{3/2}$ of $V_\pm$ can not be varied independently. Also we point out that the superpotential (\ref{e: superpotential}) is not analyzed in Refs.\  \cite{Bagchi-book}, \cite{Dutt-CES}, since we can not choose the values of the parameters $a_I$, $b_I$, $p_I$, and $\epsilon_0$ ($a_{II}$, $b_{II}$, $p_{II}$, and $q_{II}$) to simplify the superpotential $W_I$ ($W_{II}$) to our superpotential (\ref{e: superpotential}).

As in Ref.\  \cite{ALO-2015-I}, to solve the Schr\"odinger equations of the potentials (\ref{e: potentials}) we write
\begin{eqnarray}
 \frac{\dd^{2} Z_+}{\dd x^{2}} + \omega^{2}  Z_+ = \left( W^2 + \frac{\dd W}{\dd x} \right) Z_+, \nonumber \\
  \frac{\dd^{2} Z_-}{\dd x^{2}} + \omega^{2}  Z_- = \left( W^2 - \frac{\dd W}{\dd x} \right) Z_- ,
\end{eqnarray}
and note that for $\omega \neq 0$ we can transform these equations into
\begin{eqnarray}
 \left( \frac{\dd }{\dd x } + W \right) \frac{1}{i \omega} \left( \frac{\dd }{\dd x } - W \right) Z_+ = i \omega Z_+, \nonumber \\
 \left( \frac{\dd }{\dd x } - W \right) \frac{1}{i \omega} \left( \frac{\dd }{\dd x } + W \right) Z_- = i \omega Z_-,
\end{eqnarray}
from which we obtain that the functions $Z_+$ and $Z_-$ satisfy the coupled equations
\begin{eqnarray} \label{e: Z coupled}
 \left( \frac{\dd }{\dd x } - W \right) Z_+ = i \omega Z_-, \qquad  \qquad \qquad 
 \left( \frac{\dd }{\dd x } + W \right) Z_- = i \omega Z_+ .
\end{eqnarray}

Defining the functions $R_1$ and $R_2$ by $Z_+ = R_1 + R_2,$ $Z_- = R_1 - R_2,$ we find that Eqs.\ (\ref{e: Z coupled}) become
\begin{eqnarray} \label{e: tilde R}
\frac{\dd \tilde{R}_{1} }{\dd x} - i\omega \tilde{R}_{1} = i W \tilde{R}_{2}, \qquad \qquad \qquad
 \frac{\dd \tilde{R}_{2} }{\dd x} + i\omega \tilde{R}_{2} = -i W \tilde{R}_{1} , 
\end{eqnarray}
where  we take $\tilde{R}_1   =  e^{ i \pi /4} R_1 , $ $\tilde{R}_2  = e^{-i \pi /4} R_2 $. Defining the variable $z$ by\footnote{Notice that $z$ varies over the range $0 < z < 1$.}
\begin{equation} \label{e: z variable x}
 z = \frac{e^x}{e^x+1},
\end{equation} 
we get that for the superpotential (\ref{e: superpotential}) the coupled differential equations (\ref{e: tilde R}) take the form
\begin{eqnarray} \label{e: tilde R coupled}
 z(1-z) \frac{\dd \tilde{R}_{1} }{\dd z} - i \omega \tilde{R}_{1} &=& - i m z^{1/2} \tilde{R}_2 , \nonumber \\
 z(1-z) \frac{\dd \tilde{R}_{2} }{\dd z} + i \omega \tilde{R}_{2} &=&  i m z^{1/2} \tilde{R}_1 .
\end{eqnarray}

From this coupled system of ordinary differential equations in a straightforward way we obtain that the functions $\tilde{R}_{1}$ and $\tilde{R}_{2}$ are solutions of the decoupled differential equations 
\begin{align} \label{e: radial} 
\frac{\dd^{2} \tilde{R}_j }{\dd z^{2}} &+ \left( \frac{1/2}{z} - \frac{1}{1-z}\right)\frac{\dd \tilde{R}_j }{\dd z} + \frac{(i \omega/2) \epsilon }{ z^{2} (1 -z ) } \tilde{R}_j \nonumber \\  
&+ \frac{\omega^{2} }{z^{2} (1 - z)^{2}} \tilde{R}_j - \frac{m^{2}  }{z (1-z )^{2}} \tilde{R}_j  =0 ,
\end{align} 
with $j=1,2$, $\epsilon =1$ for $\tilde{R}_1$, and $\epsilon =-1$ for $\tilde{R}_2$. Taking the functions $\tilde{R}_j$ as
\begin{equation} \label{e: tilde R ansatz}
 \tilde{R}_j = z^{C_j} (1-z)^{B_j} \bar{R}_j ,
\end{equation} 
with the quantities $C_j$ and $B_j$ being solutions of the algebraic equations\footnote{We note that the parameter $\epsilon$ only appears in Eqs.\ (\ref{e: C B equations}) that determine the constants $C_1$ and $C_2$. In the equations for the constants $B_1$ and $B_2$ this parameter does not appear.}
\begin{eqnarray} \label{e: C B equations}
 & & C_j^2 - \frac{C_j}{2} + \frac{i \omega \epsilon}{2} + \omega^2 = 0,  \qquad \qquad \qquad  B_j^2 + \omega^2 - m^2 = 0, 
\end{eqnarray}
we get that the functions $\bar{R}_j$ satisfy
\begin{align} \label{e: bar R}
z(1-z) \frac{\dd^{2} \bar{R}_j }{\dd z^{2}} & + \left(2 C_j + \frac{1}{2} -  \left(2 C_j + 2 B_j + \frac{3}{2} \right) z \right)  \frac{\dd \bar{R}_j }{\dd z} \nonumber \\
& - \left(m^{2} + 2 B_j C_j + C_j + \frac{B_j}{2} - 2\omega^{2}  - \frac{i \omega \epsilon}{2}  \right) \bar{R}_j = 0 ,
\end{align}
which are hypergeometric type differential equations \cite{Abramowitz-book}--\cite{NIST-book}
\begin{equation} \label{e: hypergeometric equation}
 z(1-z)\frac{\dd^2 f}{\dd z^2} + (c -(a+b+1)z)\frac{\dd f}{\dd z} - a b f = 0 ,
\end{equation} 
with parameters
\begin{eqnarray} \label{e: a b c hypergeometric}
a_j = C_j + B_j +\tfrac{1}{2},\qquad \quad 
b_j =  C_j + B_j, \qquad \quad 
c_j = 2 C_j +\tfrac{1}{2}. 
\end{eqnarray}

Assuming that the parameters $c_j$ are not integers to discard the solutions including logarithmic terms \cite{Abramowitz-book}--\cite{NIST-book} we obtain that the functions $\bar{R}_j$ are equal to
\begin{align} \label{e: bar R hypergeometric}
 \bar{R}_j &=  G_j \,\, {}_{2}F_{1}(a_j,b_j;c_j;z) \nonumber \\
 &+ H_j \, z^{1-c_j}  {}_{2}F_{1}(a_j-c_j+1,b_j-c_j+1;2-c_j;z),
\end{align}
where ${}_{2}F_{1} (a,b;c;z)$ denotes the hypergeometric function \cite{Abramowitz-book}--\cite{NIST-book} and $G_j$, $H_j$ are constants. Therefore the functions $\tilde{R}_1$ and $\tilde{R}_2$ take the form
\begin{align} \label{e: R tilde solutions}
 \tilde{R}_j   &=  z^{C_j} (1-z)^{B_j}  [ G_j \,\, {}_{2}F_{1}(a_j,b_j;c_j;z)  \nonumber \\
    & +  H_j   \, z^{1-c_j}  {}_{2}F_{1}(a_j-c_j+1,b_j-c_j+1;2-c_j;z)  ] .
\end{align}
Considering the definitions of $\tilde{R}_j$ and $Z_\pm$, from the previous formulas we obtain the solutions of the Schr\"odinger equations (\ref{e: Schrodinger equation}) with the potentials (\ref{e: potentials}) (see Eqs.\  (\ref{e: first solution}) and (\ref{e: second solution}) below).

\section{Checking the solutions}
\label{s: verification}

Here we verify that the functions (\ref{e: R tilde solutions}) give the exact solutions to the Schr\"odinger equations of the potentials (\ref{e: potentials}). From the expressions (\ref{e: R tilde solutions}) we obtain that these functions satisfy
\begin{equation} \label{e: tilde R second derivative}
 \frac{\dd }{\dd z} \left( z(1-z) \frac{\dd \tilde{R}_j}{\dd z} \right) = \left( \frac{B_j^ 2}{1-z} + \frac{C_j^2 - C_j/2}{z}\right) \tilde{R}_j + \frac{1}{2} (1-z)\frac{\dd \tilde{R}_j}{\dd z} .
\end{equation} 
Considering that  $Z_\pm = e^{-i \pi /4} (\tilde{R}_1 \pm i \tilde{R}_2),$ from the expressions (\ref{e: C B equations}) and (\ref{e: tilde R second derivative}) we get that the functions $Z_\pm$ fulfill
\begin{equation} \label{e: Schrodinger in z variable}
 \frac{\dd }{\dd z} \left( z(1-z)\frac{\dd Z_\pm}{\dd z}  \right) + \left( \frac{\omega^2 - m^2}{1-z} + \frac{\omega^2}{z} \pm \frac{m}{2} \frac{1}{z^{1/2}} \right) Z_\pm = 0.
\end{equation} 
We can verify  that Eqs.\  (\ref{e: Schrodinger in z variable}) are the Schr\"odinger equations for the potentials (\ref{e: potentials}) but in the variable $z$. Therefore the functions $Z_\pm$ are the exact solutions to the Schr\"odinger equations of the potentials (\ref{e: potentials}).

It is convenient to notice that Eqs.\  (\ref{e: tilde R coupled}) impose some restrictions on the values of the constants $G_j$ and $H_j$. To discuss this fact in what follows we take  $B_j$ and $C_j$ in the form 
\begin{equation} \label{e: B C fixed}
 B_1 = B_2 = \sqrt{m^2 - \omega^2}, \qquad \qquad C_2 = C_1 + \tfrac{1}{2} = i \omega + \tfrac{1}{2} ,
\end{equation} 
and therefore the parameters $a_j$, $b_j$, and $c_j$ are equal to
\begin{eqnarray} \label{e: a b c fixed}
 a_2 &=& \sqrt{m^2 - \omega^2} + i \omega + 1 = b_1 + 1, \nonumber \\
 b_2 &=& \sqrt{m^2 - \omega^2} + i \omega + \tfrac{1}{2} = a_1,\\
 c_2 &=& 2 i \omega + \tfrac{3}{2} = c_1 + 1 . \nonumber
 \end{eqnarray}
Taking the solution $ \tilde{R}_1$ as
\begin{equation} \label{e: R1 tilde}
 \tilde{R}_1 = G_1 z^{C_1} (1-z)^{B_1} {}_{2}F_{1}(a_1,b_1;c_1;z),
\end{equation} 
substituting this expression in Eqs.\ (\ref{e: tilde R coupled}), we find that the function $\tilde{R}_2$ must be of the form
\begin{equation}  \label{e: R2 tilde}
  \tilde{R}_2 = \frac{i}{m} \frac{(a_1-c_1)b_1}{c_1} G_1 z^{C_2} (1-z)^{B_2} {}_{2}F_{1}(a_2,b_2;c_2;z),
\end{equation} 
that is, the constants $G_1$ and $G_2$ satisfy $ G_2 = i (a_1-c_1)b_1 G_1 / (m c_1)$.

Similar results are valid for the solution 
\begin{equation}
 \tilde{R}_1  = H_1 (1-z)^{B_1}  z^{C_1 + 1-c_1}  {}_{2}F_{1}(a_1-c_1+1,b_1-c_1+1;2-c_1;z),
\end{equation} 
since Eqs.\  (\ref{e: tilde R coupled}) produce that the function $\tilde{R}_2$ must take the form
\begin{equation}
 \tilde{R}_2 = \frac{i(1-c_1)}{m} H_1 z^{C_2}  (1-z)^{B_2} z^{1-c_2}  \hypergf  (a_2-c_2+1,b_2-c_2+1;2-c_2;z) ,
\end{equation} 
and we obtain that the constants $H_1$ and $H_2$ satisfy $H_2 =i (1-c_1)  H_1/ m $. Thus Eqs.\ (\ref{e: tilde R coupled}) impose constraints on the values of the constants $G_j$ and $H_j$.

From the previous results we get that for the Schr\"odinger equations of the potentials (\ref{e: potentials}) their linearly independent solutions  take the form
\begin{align} \label{e: first solution}
 Z_{\pm}^I & = G_1 \textrm{e}^{-i \pi / 4} [ z^{C_1} (1-z)^{B_1} \hypergf (a_1,b_1;c_1;z) \nonumber \\
 & \left. \mp \frac{(a_1-c_1)b_1}{m c_1} z^{C_2} (1-z)^{B_2} \hypergf (a_2,b_2;c_2;z) \right],
\end{align} 
 and
\begin{align}\label{e: second solution}
 Z_{\pm}^{II} & = H_1 \textrm{e}^{-i \pi / 4} [ z^{C_1+1-c_1} (1-z)^{B_1}  \hypergf (a_1-c_1+1,b_1-c_1+1;2-c_1;z) \nonumber \\
 & \left. \mp \frac{1-c_1}{m} z^{C_2+1-c_2} (1-z)^{B_2}   \hypergf  (a_2-c_2+1,b_2-c_2+1;2-c_2;z) \right].
\end{align}
It is convenient to notice that using the recurrence relations for the contiguous hypergeometric functions \cite{Lebedev} and Eqs.\ (\ref{e: tilde R}) we can show in a direct way that the functions $Z_{\pm}^I$ and $Z_{\pm}^{II}$ are solutions of the Schr\"odinger equations for the potentials (\ref{e: potentials}).

Furthermore, taking into account that the Wronskian of the linearly independent solutions to the hypergeometric equation (\ref{e: hypergeometric equation}) is \cite{NIST-book}
\begin{equation}
 \mathcal{W}_z [ \hypergf (a,b;c;z), z^{1-c} \hypergf (a-c+1, b-c+1; 2-c;z) ] = (1-c)z^{-c} (1-z)^{c-a-b-1},
\end{equation} 
we find that the Wronskians of the solutions (\ref{e: first solution}) and (\ref{e: second solution}) are equal to (for $G_1=H_1=1$)
\begin{equation}
  \mathcal{W}_z [Z_{\pm}^I,Z_{\pm}^{II} ] = \pm \frac{2 \omega}{m} (1-c_1) .
\end{equation} 
As expected, they are constants.


\section{Scattering}
\label{s: scattering}

As we commented previously, the potentials (\ref{e: potentials}) remind us to the step potential (see Fig.\ \ref{figure1}) and therefore we believe that the potentials $V_\pm$ are appropriate to discuss scattering problems. Thus in what follows we calculate the reflection and transmission amplitudes of the partner potentials $V_\pm$. We first discuss the potential $V_+$ and then the potential $V_-$. 

To begin, we notice that in the limits $x \to \pm \infty$ the variable $z$ defined in the formula (\ref{e: z variable x}) behaves as
\begin{eqnarray}
 & \lim_{x \to +\infty} z \approx 1, \qquad  \qquad & \lim_{x \to +\infty}(1- z) \approx \textrm{e}^{-x} , \nonumber  \\
 & \lim_{x \to -\infty} z \approx \textrm{e}^{x},  \qquad  \qquad & \lim_{x \to -\infty}(1- z) \approx 1 .
\end{eqnarray}
From the expressions (\ref{e: first solution}) and (\ref{e: second solution}) for the solutions $Z_+^I$ and $Z_+^{II}$ of the potential $V_+$ in the limit $x \to - \infty$ we obtain that they behave as
\begin{equation} \label{e: Z asymptotic minus}
 \lim_{x \to -\infty} Z_+^I \approx G_1 \textrm{e}^{- i \pi / 4}   \textrm{e}^{i \omega x}, \qquad  \lim_{x \to -\infty} Z_+^{II} \approx H_1 \textrm{e}^{- i \pi / 4} \left( \frac{c_1-1}{m} \right)  \textrm{e}^{- i \omega x}.
\end{equation} 
Taking into account Kummer's formula for the hypergeometric function \cite{Abramowitz-book}--\cite{NIST-book}, 
\begin{eqnarray} \label{e: Kummer property y 1-y}
& &{}_2F_1(a,b;c;x) = \frac{\Gamma(c) \Gamma(c-a-b)}{\Gamma(c-a) \Gamma(c - b)} {}_2 F_1 (a,b;a +b +1-c;1-x)  \\
&+& \frac{\Gamma(c) \Gamma( a + b - c)}{\Gamma(a) \Gamma(b)} (1-x)^{c-a-b} {}_2F_1(c-a, c-b; c + 1 -a-b; 1 -x). \nonumber
\end{eqnarray}
we get that as $x \to +\infty$ the functions $Z_+^I$ and $Z_+^{II}$ behave in the form
\begin{eqnarray} \label{e: Z asymptotic plus}
 \lim_{x \to +\infty} Z_+^I & \approx & G_1 \textrm{e}^{- i \pi / 4} \left[  \textrm{e}^{- i \sqrt{\omega^2 - m^2 } x} \frac{\Gamma(c_1) \Gamma(c_1-a_1-b_1)}{\Gamma(c_1-a_1)\Gamma(c_1-b_1)} \left( 1+ \frac{b_1}{m} \right) \right. \nonumber  \\
 & + & \left.  \textrm{e}^{ i \sqrt{\omega^2 - m^2 } x} \frac{\Gamma(c_1) \Gamma(a_1+b_1-c_1)}{\Gamma(a_1)\Gamma(b_1)} \left( 1 + \frac{c_1-a_1}{m}\right) \right] ,   \\
 \lim_{x \to +\infty} Z_+^{II} & \approx &  H_1 \textrm{e}^{- i \pi / 4} \left[  \textrm{e}^{- i \sqrt{\omega^2 - m^2 } x} \frac{\Gamma(2-c_1) \Gamma(c_1-a_1-b_1)}{\Gamma(1-a_1)\Gamma(1-b_1)} \left( 1+ \frac{b_1}{m} \right) \right. \nonumber  \\
 & + & \left.  \textrm{e}^{ i \sqrt{\omega^2 - m^2 } x} \frac{\Gamma(2-c_1) \Gamma(a_1+b_1-c_1)}{\Gamma(a_1-c_1+1)\Gamma(b_1-c_1+1)} \left( 1 + \frac{c_1-a_1}{m}\right) \nonumber \right] . 
\end{eqnarray} 
In the previous formulas and in what follows we assume that $\omega > m$.

To solve the scattering problem for the potentials $V_\pm$ we impose that the solutions of the Schr\"odinger equation (\ref{e: Schrodinger equation}) must behave in the form
\begin{equation} \label{e: assumed behavior minus}
 Z_\pm \approx \textrm{e}^{i \omega x} + R^\pm \textrm{e}^{- i \omega x}
\end{equation} 
as $x \to - \infty$ and in the form
\begin{equation} \label{e: assumed behavior plus}
 Z_\pm \approx T^\pm  \textrm{e}^{ i \sqrt{\omega^2 - m^2 } x} 
\end{equation} 
as  $x \to + \infty$. In these formulas and in what follows $ R^+$ and $ T^+$ ($ R^-$ and $ T^-$) denote the reflection and transmission amplitudes of the potential $V_+$ ($V_-$).

As is well known \cite{Dterhaar}, if the solution of the Schr\"odinger equation takes the form
\begin{equation}
 Z = E_I F^I + E_{II} F^{II} ,
\end{equation} 
where $E_I$ and $E_{II}$ are constants, and the functions $F^\kappa$ behave as 
\begin{eqnarray}
 \lim_{x \to +\infty} F^\kappa \approx a^{\kappa +} \textrm{e}^{i \omega x} + b^{\kappa +} \textrm{e}^{- i \omega x} ,\nonumber \\
 \lim_{x \to -\infty} F^\kappa \approx a^{\kappa -} \textrm{e}^{i \omega x} + b^{\kappa -} \textrm{e}^{- i \omega x} ,
\end{eqnarray} 
where $ a^{\kappa +}$, $b ^{\kappa +}$, $a^{\kappa -}$,  $b^{\kappa -}$ are constants, and $\kappa =I,II$, then the reflection and transmission amplitudes are  \cite{Khare-scattering}, \cite{Dterhaar}
\begin{equation}
 R = \frac{b^{II +} b^{I -} -   b^{I +} b^{II -} }{ a^{I -} b^{II +} - b^{I +} a^{II -}  }, \qquad \quad T = \frac{b^{II +} a^{I +} -  b^{I +} a^{II +}   }{ a^{I -} b^{II +} - b^{I +} a^{II -} }.
\end{equation} 

For the solutions $Z^I_+$ and $Z^{II}_+$, from the formulas (\ref{e: Z asymptotic minus}) and (\ref{e: Z asymptotic plus}) we deduce that the quantities $ a^{\kappa +}$, $b ^{\kappa +}$, $a^{\kappa -}$,  $b^{\kappa -}$ are
\begin{eqnarray}
a^{I -} &=& 1, \qquad \qquad \qquad a^{II -} = 0,  \nonumber \\
b^{I -} &=& 0, \qquad \qquad \qquad b^{II -}  = \frac{c_1-1}{m} , \nonumber \\
a^{I +} &=& \frac{\Gamma(c_1) \Gamma(a_1+b_1-c_1)}{\Gamma(a_1)\Gamma(b_1)} \left( 1 + \frac{c_1-a_1}{m}\right),   \nonumber \\
a^{II +} &=& \frac{\Gamma(2-c_1) \Gamma(a_1+b_1-c_1)}{\Gamma(a_1-c_1+1)\Gamma(b_1-c_1+1)} \left( 1 + \frac{c_1-a_1}{m}\right),\\
b^{I +}  &=& \frac{\Gamma(c_1) \Gamma(c_1-a_1-b_1)}{\Gamma(c_1-a_1)\Gamma(c_1-b_1)} \left( 1+ \frac{b_1}{m} \right) , \nonumber \\
b^{II +}  &=&  \frac{\Gamma(2-c_1) \Gamma(c_1-a_1-b_1)}{\Gamma(1-a_1)\Gamma(1-b_1)} \left( 1+ \frac{b_1}{m} \right). \nonumber
\end{eqnarray} 
Taking into account the previous expressions for the potential $V_+$ we obtain that its reflection and transmission amplitudes are equal to
\begin{eqnarray} \label{e: reflection transmission V+}
 R^+ &=& - \frac{b^{II -} b^{I +} }{b^{II +}} = \frac{1}{m} \frac{\Gamma(1-a_1) \Gamma(1-b_1) \Gamma(c_1) }{\Gamma(1-c_1) \Gamma(c_1-a_1)\Gamma(c_1-b_1)}, \\
 T^+ &=& \frac{b^{II +} a^{I +} - b^{I +} a^{II +} }{b^{II +}} =  \left( 1 + \frac{c_1-a_1}{m}\right) \frac{\Gamma(1-a_1) \Gamma(1-b_1)}{\Gamma(1-c_1) \Gamma(1 + c_1-a_1-b_1) } .  \nonumber
\end{eqnarray}
From these expressions we find that\footnote{The asterisk denotes the complex conjugation.}
\begin{eqnarray} \label{e: reflection transmission coefficients}
 R^+ R^{+ *} &=& \frac{\sinh (2 \pi \omega - 2 \pi \sqrt{\omega^2 - m^2})}{\sinh (2 \pi \omega + 2 \pi \sqrt{\omega^2 - m^2})} ,  \\
 T^+ T^{+ *} &=& \frac{\omega}{\sqrt{\omega^2 - m^2}} \frac{ 2 \cosh (2 \pi \omega ) \sinh (2 \pi \sqrt{\omega^2 - m^2})}{\sinh (2 \pi \omega + 2 \pi \sqrt{\omega^2 - m^2})} . \nonumber 
\end{eqnarray} 

Furthermore, it is well known that for the Schr\"odinger equation (\ref{e: Schrodinger equation}) the quantity 
\begin{equation}
 Z^* \frac{\dd Z}{\dd x} - Z \frac{\dd Z^*}{\dd x} ,
\end{equation} 
is a constant \cite{Flugge}, \cite{Dterhaar}. For the approximate solutions (\ref{e: assumed behavior minus}) and (\ref{e: assumed behavior plus}) we get that the previous condition implies that the reflection and transmission amplitudes satisfy 
\begin{equation} \label{e: flux conserved}
 R^+ R^{+ *} + \frac{\sqrt{\omega^2 - m^2} }{\omega} T^+ T^{+ *} = 1 .
\end{equation} 
An straightforward calculation shows that the expressions (\ref{e: reflection transmission V+}) for the reflection and transmission amplitudes of the potential $V_+$ fulfill the condition (\ref{e: flux conserved}).

\begin{figure}[th]
\begin{center}
\includegraphics[scale=.9,clip=true]{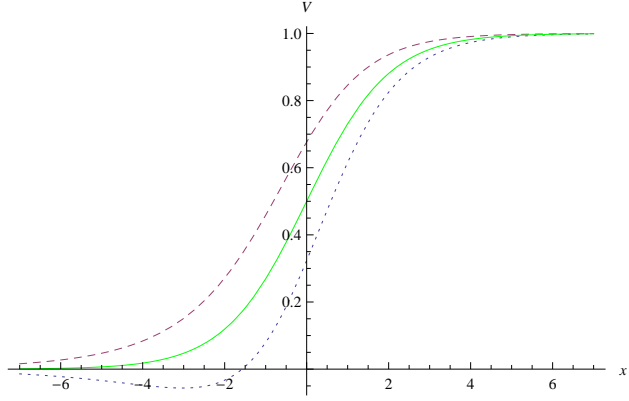}
\caption{Plots of the potentials $V_+$ (dotted line), $V_-$ (dashed line) and $V_S$ (solid line) for $V_0=m^2$, $\alpha =1$, and $m=1$. \label{figure3}} 
\end{center}
\end{figure}

A similar computation shows that for the potential $V_-$ its reflection and transmission amplitudes take the form
\begin{eqnarray} \label{e: relationship minus plus}
 R^- &=& - \frac{1}{m} \frac{\Gamma(1-a_1) \Gamma(1-b_1) \Gamma(c_1) }{\Gamma(1-c_1) \Gamma(c_1-a_1)\Gamma(c_1-b_1)} = -R^+,  \\
 T^-  &=& \left( 1 - \frac{c_1-a_1}{m} \right) \frac{\Gamma(1-a_1) \Gamma(1-b_1)}{\Gamma(1-c_1) \Gamma(1 + c_1-a_1-b_1) } = \frac{\left( 1 - \tfrac{c_1-a_1}{m}\right)}{\left( 1 + \tfrac{c_1-a_1}{m}\right)}  T^+ . \nonumber 
\end{eqnarray} 
Simplifying the last expression we also find the relationship
\begin{equation}  \label{e: relationship transmission}
 T^+ = \frac{i \omega}{m + i \sqrt{\omega^2 - m^2}} T^- .
\end{equation} 
From these expressions for the amplitudes $R^-$ and $T^-$ we obtain that $R^- R^{- *} = R^+ R^{+ *}$ and $T^- T^{- *} = T^+ T^{+ *}$, that is, the amplitudes $R^-$ and $T^-$ of the potential $V_-$ satisfy the condition (\ref{e: flux conserved}).

As we commented before, the potentials $V_\pm$ are partner potentials. Thus SUSYQM predicts a relationship between their reflection (transmission) amplitudes  (see for example the formulas (16.7) and (16.8) of Ref.\  \cite{Gangopadhyaya book}). From these expressions for the potentials $V_\pm$ we get
\begin{equation} \label{e: SUSY reflection transmission}
 R^+ = -R^-, \qquad \qquad T^+ = \frac{ik}{m+ik^\prime} T^- ,
\end{equation} 
where the quantities $k$  and $k^\prime$ are equal to 
\begin{equation}
 k= \sqrt{\omega^2 - W_-^2}= \omega, \qquad k^\prime= \sqrt{\omega^2 - W_+^2}=\sqrt{\omega^2-m^2} .
\end{equation} 
We see that the formulas (\ref{e: SUSY reflection transmission}) coincide with the previously obtained relationships (\ref{e: relationship minus plus}) and (\ref{e: relationship transmission}). Notice that SUSYQM does not allow to calculate the full expression for the reflection and transmission amplitudes (\ref{e: reflection transmission V+}) and (\ref{e: relationship minus plus}).

\begin{figure}[th]
\begin{center}
\includegraphics[scale=.9,clip=true]{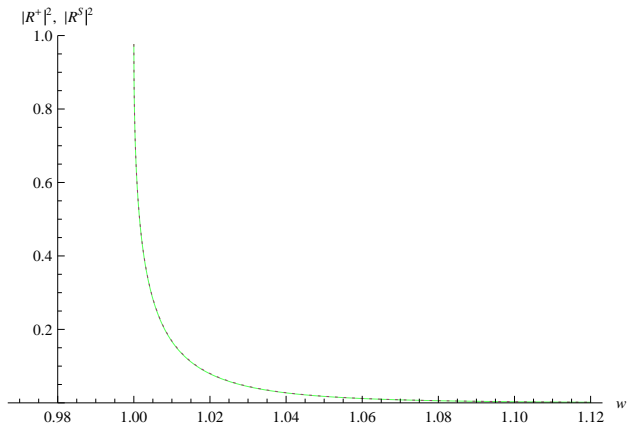}
\caption{Plots of the reflection coefficients for $\omega > m$ of the potential $V_+$  (dotted line) and for the step potential $V_S$ (solid line). We take $V_0=m^2$, $\alpha =1$, and $m=1$. \label{figure4}} 
\end{center}
\end{figure}

\begin{figure}[th]
\begin{center}
\includegraphics[scale=.9,clip=true]{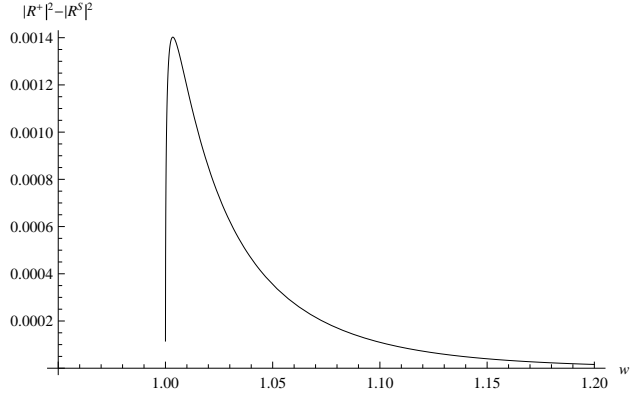}
\caption{Difference of the reflection coefficients for $\omega > m$ of the potentials $V_+$ and $V_S$. In this plot we take $V_0=m^2$, $\alpha =1$, and $m=1$. \label{figure5}} 
\end{center}
\end{figure}

We think that the partner potentials $V_\pm$ generalize the potential step (\ref{e: potential step}) and we observe that their shapes are similar (see Fig.\ \ref{figure3}). Since the reflection amplitude of the step potential  (\ref{e: potential step}) is equal to (in what follows for the potential step we take $V_0=m^2$, $\alpha=1$) \cite{Flugge}
\begin{equation}
 R^S = \frac{\Gamma(2 i \omega) \Gamma(-i \omega - i\sqrt{\omega^2-m^2})  \Gamma(1 -i \omega - i\sqrt{\omega^2-m^2}) }{\Gamma(- 2 i \omega) \Gamma(i \omega - i\sqrt{\omega^2-m^2}) \Gamma(1 + i \omega - i\sqrt{\omega^2-m^2})} ,
\end{equation} 
and therefore we get that its reflection coefficient takes the form 
\begin{equation} \label{e: step reflection coefficient}
 R^S R^{S *} = \frac{\sinh^2(\pi \omega - \pi \sqrt{\omega^2-m^2}) }{\sinh^2(\pi \omega + \pi \sqrt{\omega^2-m^2})} .
\end{equation} 
In Fig.\ \ref{figure4} we compare the reflection coefficient (\ref{e: reflection transmission coefficients}) of the potential $V_+$ and the reflection coefficient (\ref{e: step reflection coefficient}) of the potential step (\ref{e: potential step}). We see that apparently their reflection coefficients are equal, but plotting their difference (see Fig.\ \ref{figure5}) we notice that they are slightly different, but this difference is not visible in Fig.\ \ref{figure4}.

As is well known, the potential step (\ref{e: potential step}) does not have transmission resonances  \cite{Boonserm-Visser}. In a similar way from the expressions (\ref{e: reflection transmission V+}) and (\ref{e: relationship minus plus}) we obtain that the potentials $V_\pm$ do not have transmission resonances. Closely related to the transmission resonances are the quasinormal frequencies, defined as the complex frequencies corresponding to purely outgoing waves as $x \to \pm \infty$ \cite{Boonserm-Visser}. Thus we find the quasinormal frequencies by looking for the complex numbers where the transmission amplitude becomes infinite \cite{Boonserm-Visser}. The transmission amplitude for the potential step (\ref{e: potential step}) is equal to \cite{Flugge}, \cite{Boonserm-Visser}
\begin{equation}
 T^S = \frac{\Gamma(1 -i \omega - i\sqrt{\omega^2-m^2}) \Gamma(-i \omega - i\sqrt{\omega^2-m^2})  }{\Gamma(- 2 i \omega)  \Gamma(1 - 2 i \sqrt{\omega^2-m^2}) } .
\end{equation} 
Therefore its quasinormal frequencies are determined by the poles of the numerator of the amplitude $T^S$ located at \cite{Boonserm-Visser}
\begin{equation}
 1 - i \omega - i\sqrt{\omega^2-m^2} = -n,
\end{equation}
with $n=0,1,\dots$ Thus its quasinormal frequencies are equal to \cite{Boonserm-Visser}
\begin{equation} \label{e: quasinormal step potential}
 \omega_S = - \frac{i}{2} \left( n+1 - \frac{m^2}{n+1}\right).
\end{equation} 

In a similar way, from the expression (\ref{e: reflection transmission V+}) for the transmission amplitude $T^+$ of the potential $V_+$ we find that it transforms into 
\begin{equation}
 T^+ = \left( 1 + \frac{c_1-a_1}{m} \right) \frac{\sqrt{\pi}}{2^{-2 i \omega - 2 i \sqrt{\omega^2 -m^2}}} \frac{\Gamma(1 -2 i \omega -2 i \sqrt{\omega^2-m^2})}{\Gamma(\tfrac{1}{2} - 2 i \omega)   \Gamma(1 - 2 i \sqrt{\omega^2-m^2})} .
\end{equation} 
We notice that it becomes infinite at the poles of the numerator, that is, the quasinormal frequencies of the potential $V_+$ are determined by
\begin{equation}
 1 -2 i \omega -2 i \sqrt{\omega^2-m^2} = -n, \qquad \qquad n=0,1,2,\dots
\end{equation} 
Thus the quasinormal frequencies of the potential $V_+$ are equal to
\begin{equation} \label{e: quasinormal V plus}
 \omega_+ = - \frac{i}{4} \left( n+1 - \frac{4 m^2}{n+1}\right).
\end{equation} 
A similar result is valid for the potential $V_-$. Hence we must add the potentials $V_\pm$ to the list of Ref.\  \cite{Boonserm-Visser} that enumerates the potentials with exactly calculated quasinormal frequencies.

Comparing the quasinormal frequencies (\ref{e: quasinormal step potential}) and (\ref{e: quasinormal V plus}) for the potential step and the potential $V_+$, we find that their mathematical forms are similar, but there are relevant differences. For example, in the asymptotic limit $n \to \infty$ the quasinormal frequencies (\ref{e: quasinormal step potential}) behave as
\begin{equation}
 \omega_S \approx - \frac{i}{2} n ,
\end{equation} 
whereas the quasinormal frequencies (\ref{e: quasinormal V plus}) behave in the form
\begin{equation}
 \omega_+ \approx - \frac{i}{4} n .
\end{equation} 
Furthermore  in the quasinormal frequencies (\ref{e: quasinormal step potential}) and (\ref{e: quasinormal V plus}) the factor multiplying the parameter $m$ is different.


\section{Final remarks}
\label{s: remarks}

To finish this work we note the following facts about the potentials (\ref{e: potentials}).

\begin{enumerate}

\item From the expressions (\ref{e: potentials}) for the potentials $V_+$ and $V_-$ we notice that they fulfill $V_+(x,m) = V_-(x,-m)$. If $\hat{\alpha}_0$,  $\hat{\alpha}_1$ are parameters, $R_0 (\hat{\alpha}_0)$ is a function of $\hat{\alpha}_0$ and comparing the previous expressions  with the formula $ V_+(x,\hat{\alpha}_0) = V_-(x,\hat{\alpha}_1) + R_0 (\hat{\alpha}_0) ,$ that defines the shape invariance condition for the partner potentials \cite{Gendenshtein}, we find that for the potentials (\ref{e: potentials}), the quantities $\hat{\alpha}_0$,  $\hat{\alpha}_1$, and   $R_0 (\hat{\alpha}_0)$ are equal to $\hat{\alpha}_0 = m,$ $\hat{\alpha}_1= -m = - \hat{\alpha}_0 ,$ $R_0 (\hat{\alpha}_0) = 0,$ that is $ \hat{\alpha}_1= q \hat{\alpha}_0$ with $q=-1$. Therefore the potentials $V_+$ and $V_-$ of the formulas (\ref{e: potentials}) are multiplicative shape invariant \cite{Cooper-book}--\cite{Bagchi-book}. Notice that we obtain the potentials (\ref{e: potentials}) in closed form, in contrast, the previously known multiplicative shape invariant potentials that include a scaling of the parameters are given in series form \cite{Cooper-book}.

The previous formula $\hat{\alpha}_1= - \hat{\alpha}_0$ remind us a reflection, but on the parameters of the potential, which is different from the recent analyzed shape invariance with reflection transformations of Ref.\  \cite{Aleixo-Balantekin}, where they consider reflections of the coordinate and translations of the parameters.

\item We note that the indefinite integrals of the potentials (\ref{e: potentials}) are equal to
\begin{equation}
 \int V_\pm (x) \dd x = m^2 \ln (e^x +1) \mp m \frac{e^{x/2}}{(e^{x} + 1)^{1/2}} + C_\pm ,
\end{equation} 
where $C_\pm $ are constants and the integrals 
\begin{equation}
 \int_{-\infty}^{+\infty} V_\pm (x) \dd x
\end{equation} 
diverge to $+ \infty$.

\item Furthermore we notice that for the Schr\"odinger equations of the potentials (\ref{e: potentials}) their linearly independent solutions include two hypergeometric functions with non constant coefficients  (see the formulas (\ref{e: first solution}) and (\ref{e: second solution})), in a similar way to the exact solutions studied previously in Refs.\  \cite{ALO-2015-I}, \cite{Ishkhanyan-1}, \cite{Ishkhanyan-2}, but in the last references are involved confluent hypergeometric functions. We have not found a transformation that simplifies this sum of hypergeometric functions to a single hypergeometric function, but this problem must be analyzed carefully. Thus if the previous transformation is not possible, the potentials (\ref{e: potentials}) are an example of those proposed but not given in explicit form in Ref.\ \cite{Cooper:1986tz} whose linearly independent solutions include a sum of hypergeometric functions.

Undoubtedly we must study the use of the potentials $V_\pm$ as a basis to generate new exactly solvable potentials \cite{Cooper-book}.

\end{enumerate}

\section{Acknowledgments}

We thank the support by CONACYT M\'exico, SNI M\'exico, EDI-IPN, COFAA-IPN, and Research Projects IPN SIP-20150707 and IPN SIP-20151031.

\end{document}